\documentclass[conference]{IEEEtran}
\IEEEoverridecommandlockouts

\usepackage{cite}
\usepackage{amsmath,amssymb,amsfonts}
\usepackage{algorithmic}
\usepackage{textcomp}
\usepackage[table]{xcolor}
\usepackage{graphicx}
\usepackage{url}
\usepackage{booktabs}
\usepackage{caption}
\usepackage{multirow}
\usepackage{array}
\usepackage{placeins}
\usepackage{stfloats}
\usepackage{pifont}
\usepackage{listings}
\usepackage{balance}
\usepackage{cmap}
\usepackage{threeparttable}
\usepackage{colortbl}
\usepackage{comment}
\usepackage[inline]{enumitem}
\usepackage{tabularx}
\usepackage[most]{tcolorbox}
\usepackage{hyperref}

\hypersetup{colorlinks=false, hidelinks}
\definecolor{paleblue}{rgb}{0.85,0.9,1.0}
\definecolor{deepblue}{rgb}{0.05,0.35,0.75}

\tcbset{
  mybluebox/.style={
    enhanced,
    colback=paleblue,
    frame hidden,
    borderline west={2.5pt}{0pt}{deepblue},
    boxsep=1ex,
    sharp corners
  }
}

\newcolumntype{Y}{>{\raggedright\arraybackslash}X}

\begin{document}

\title{ToMAS: A Pilot Failure-Grounded Theory-of-Mind Benchmark from Multi-Agent LLM Failures}
\author{
\IEEEauthorblockN{Muhammad Ashar Ishfaq\IEEEauthorrefmark{1} and
Glaucia Melo\IEEEauthorrefmark{2}}
\IEEEauthorblockA{\IEEEauthorrefmark{1}\textit{The Islamia University of Bahawalpur}, Pakistan \\
s24barin1m01132@iub.edu.pk}
\IEEEauthorblockA{\IEEEauthorrefmark{2}\textit{Toronto Metropolitan University}, Toronto, Canada \\
glaucia@torontomu.ca}
}

\maketitle

\begin{abstract}
LLM-based multi-agent systems can fail even when communication succeeds because agents do not correctly track their peers' roles, knowledge, or intentions.
We investigate whether such inter-agent misalignment cases, labelled FC2 in MAST-Data, can be converted into functional partner-state reasoning items.
ToMAS applies four explicit convertibility criteria to diagnosed execution traces.
A full conversion pass over 242 eligible non-AG2 training traces produced 39 \textsc{Clean} items.
In an 18-trace reliability pilot, two annotators achieved 94.4\% raw agreement and Cohen's $\kappa=0.92$.
We then used the converted items as binary rewards in a small-scale GRPO feasibility experiment with Qwen2.5-1.5B.
On a 28-item held-out Magentic GAIA diagnostic, every evaluated condition exceeded the ROUGE-L threshold on the same 2 of 28 items. Post-hoc adapter checks show why: under the learning rate used, the LoRA update remained numerically negligible ($\max|\Delta W|\approx 7{\times}10^{-6}$), so all conditions decode identically to the untrained checkpoint.
The experiment therefore does not show a training effect and cannot establish one; it reports an executable pipeline together with two limitations that any conclusive study must address: a provenance gap between the training and evaluation items, and lexical-overlap scoring. ToMAS provides a preliminary rubric and pipeline for converting diagnosed coordination failures into trainable partner-state reasoning items and identifies the requirements for a conclusive matched-domain evaluation.
\end{abstract}


\section{Introduction}

LLM-based multi-agent systems distribute complex work across specialized agents---for software engineering~\cite{hong2024metagpt,qian2024chatdev}, mathematical reasoning~\cite{wu2024autogen}, and general-purpose automation~\cite{fourney2024magentic}.
The premise is that role specialization enables decomposition, cross-checking, and stronger collective performance than a single agent.
Empirically, coordination frequently breaks down when agents hold different knowledge and responsibilities.
Analysis of 1,642 annotated execution traces across seven frameworks identifies inter-agent misalignment as a persistent failure mode that survives structural fixes to other coordination surfaces~\cite{cemri2025mast}.

Coordination failures in multi-agent LLM systems have drawn two largely separate research responses.
Failure taxonomies characterize when coordination breaks down---agents misalign on roles, knowledge, or intentions---from annotated execution traces or controlled fault injection~\cite{cemri2025mast,jia2026masfire,zhang2026silobench}.
ToM benchmarks and RL methods train social reasoning for coordination in abstract games, embodied households, and cooperative MARL environments~\cite{agashe2025llmcoord,juneja2026enacttom,tombroken2025,cai2026maro,yao2026langmarl,li2025tom}.
Diagnostic work typically stops at recommending system changes; ToM and RL training typically use authored tasks rather than FC2 failures from observed execution traces~\cite{cemri2025mast,juneja2026enacttom,yao2026langmarl}.
None uses MAST-Data FC2 cases as the joint source for benchmark items and RL reward signals~\cite{cemri2025mast}.
Diagnosed coordination failures seldom become the explicit objects against which agents are trained and evaluated.

Selected diagnosed FC2 failures can serve as supervision rather than only as explanation.
The subset of MAST FC2 failures that can be represented as functional partner-state reasoning problems can supply benchmark items and RL rewards from the same source.
FC2 refers to inter-agent misalignment: cases where agents fail to coordinate because they do not correctly track, interpret, or respond to other agents' roles, knowledge, or intentions.
FC2 is broader than Theory of Mind; other FC2 labels, such as tool-loop or protocol failures without a peer belief or knowledge target, cannot be represented as partner-state items.
This paper investigates two research questions:

\begin{enumerate}
    \item \textbf{RQ1:} Which MAST FC2 inter-agent misalignment
    traces can be converted into functional partner-state reasoning
    items under explicit inclusion criteria, and how does
    convertibility vary across the available frameworks and model
    outputs?

    \item \textbf{RQ2:} What does a small-scale GRPO feasibility
    experiment reveal about the use of converted FC2 items as binary
    reward signals and their generalization to a separately sourced
    evaluation domain?
\end{enumerate}

We introduce ToMAS, a pilot benchmark-construction and training
pipeline that filters FC2-labeled traces from MAST-Data through four
explicit convertibility criteria. ToMAS converts each retained trace
into an item asking what a peer knew and what the acting agent should
therefore have done. The paper makes three contributions:

\begin{itemize}
    \item a four-criterion rubric for determining whether an FC2
    trace contains an isolable functional partner-state reasoning
    problem;

    \item an empirical conversion analysis of 242 eligible non-AG2
    training traces, producing 39 confirmed \textsc{Clean} items; and

        \item a small-scale GRPO diagnostic showing that the resulting
    reward pipeline can be executed end to end, together with a
    transparent post-hoc account of why this run cannot support a
    training claim: the optimization left the policy effectively
    unchanged, so the held-out comparison is uninformative about
    transfer.
\end{itemize}

The purpose of this pilot is therefore to establish the
feasibility and boundaries of failure-grounded benchmark
construction. We do not claim that ToMAS solves multi-agent
coordination or that the present experiment demonstrates a reduction
in FC2 failures. Establishing such an effect requires a larger
benchmark, a matched-domain held-out evaluation, and correctness
measures that go beyond lexical overlap.

\section{Related Work}

\subsection{Multi-agent complexity}

Multi-agent LLM systems are built on the premise that dividing work across specialized roles lets a system decompose tasks and cross-check its own outputs~\cite{hong2024metagpt,qian2024chatdev,wu2024autogen,fourney2024magentic}.
Under matched cost and compute, the simpler baseline generally wins~\cite{xia2025agentless,kapoor2024agents,elites2026}.
Empirically, this promise has not materialized at scale.
A simple agentless pipeline achieves among the highest open-source performance on SWE-Bench Lite at lower cost~\cite{xia2025agentless}.
Best-of-$N$ sampling frequently matches or outperforms multi-agent designs under equalized budget~\cite{kapoor2024agents}.
Adding agents can degrade outcomes once coordination overhead compounds~\cite{elites2026}.
These results do not merely recommend fewer agents.
They point at coordination---not model capability---as the suspect: agents still fail when knowledge and responsibility are distributed~\cite{cemri2025mast}.
Understanding \emph{why} coordination breaks, not merely that it does, is the precondition for any training intervention.

\subsection{Diagnosis: coordination as the bottleneck}

Systematic diagnosis has since located that bottleneck.
A 14-mode taxonomy derived from 1,642 annotated execution traces across seven frameworks shows that most failures trace back to specification and coordination defects rather than base-model weaknesses~\cite {cemri2025mast}.
Within this taxonomy, FC2 denotes inter-agent misalignment: cases where agents fail to coordinate because they do not correctly track, interpret, or respond to other agents' roles, knowledge, or intentions.
Structural interventions do work on other surfaces.
Closed-loop architectures neutralize over 40\% of faults that cause catastrophic collapse in linear pipelines~\cite{jia2026masfire}.
Encoded role specifications and SOPs reduce design-level failures~\cite{hong2024metagpt,qian2024chatdev}.
Yet inter-agent misalignment survives them.
Even when agents successfully exchange messages, they often fail to synthesize distributed information.
The breakdown is localized to reasoning integration rather than message exchange~\cite{zhang2026silobench}.
Diagnosis therefore names the residue.
Diagnosis has not, so far, been trained against it.

\subsection{Partner-state reasoning, and the literal/functional split}

One candidate explanation for that residue is a capability gap rather than a design gap.
Models score above 80\% on comprehension of a coordination environment while falling below 40\% on questions requiring them to model a partner's knowledge~\cite{agashe2025llmcoord,lamalfa2025mark}.
Strong literal belief prediction does not imply functional use of that belief when choosing an action~\cite{tombroken2025,zerocost2026}.
We do not equate all FC2 with ToM.
We hypothesize that \emph{some} FC2 failures reflect deficits in functional partner-state reasoning.
Other FC2 cases may be protocol, tool-use, or design issues without a peer belief or knowledge target.
Frontier models can report partner beliefs and still score 0\% on hard functional ToM splits in embodied coordination~\cite{juneja2026enacttom}.
Related work on zero-cost collaboration shows that capable models can still coordinate poorly when helping is free~\cite{zerocost2026}.
This is a cooperation deficit distinct from FC2 misalignment under distributed knowledge.
This evidence comes from abstract games, isolated probes, or embodied households.
The evidence does not come from the convertible subset of MAS execution failures documented in diagnostic research~\cite{cemri2025mast}.
The subset qualifier is earned here: we do not treat every FC2 label as a ToM failure.

\subsection{Training social reasoning: a toolkit with an authored reward}

Reinforcement learning has become the standard route to capabilities that supervised fine-tuning does not reach~\cite{yu2025dapo}.
Applied to social reasoning, it improves cooperation through step-wise credit assignment~\cite{cai2026maro}, embodied functional-ToM objectives~\cite{juneja2026enacttom}, language-level credit assignment~\cite{yao2026langmarl}, and explicit partner modeling before action selection~\cite{li2025tom}.
These methods provide useful training tools.
Their rewards are authored from constructed environments rather than observed MAS execution failures.
ToMAS treats selected labelled failures as supervision.
ToMAS converts cases with an identifiable peer state into benchmark items and RL rewards, under explicit inclusion criteria~\cite{cemri2025mast}.
Structural repair and generic social-reasoning training remain necessary for other failure modes.

\section{Methodology}

\begin{figure}[t]
    \centering
    \includegraphics[
        width=\columnwidth,
        draft=false
    ]{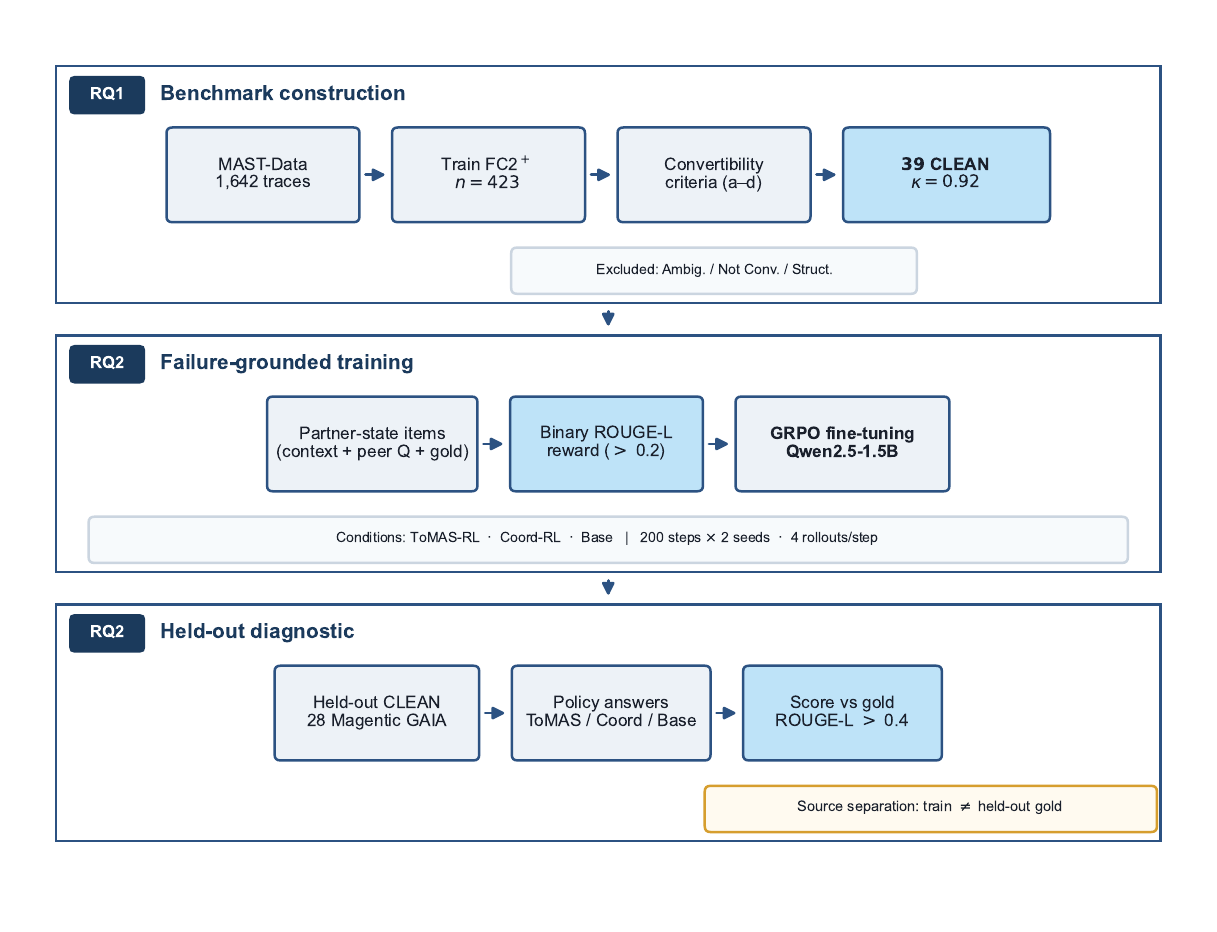}
    \vspace{2pt}
    \caption{ToMAS workflow. Benchmark construction filters
    FC2$^{+}$ training traces into 39 \textsc{Clean} partner-state
    items (RQ1). The items are then used as binary ROUGE-L rewards in
    a small-scale GRPO experiment. A separately sourced held-out pool
    is used to diagnose the limitations of the present training and evaluation setup
    (RQ2).}
    \label{fig:workflow}
\end{figure}

We organize the methodology around benchmark construction (RQ1) and
the exploratory training diagnostic (RQ2). Source traces and
convertibility criteria determine which FC2 failures can produce
partner-state items. Item construction and reward specification
define the ToMAS training signal. The training conditions and
held-out evaluation assess whether the current experimental design
can support a comparative training claim. Figure \ref{fig:workflow} presents an overview of our methodology. 
\subsection{Source traces}

We use the public Hugging Face release of MAST-Data~\cite{cemri2025mast} (\texttt{mcemri/MAST-Data}, accessed 27 August 2026), which contains 1,642 annotated traces across seven MAS frameworks (AG2, ChatDev, MetaGPT, AppWorld, HyperAgent, Magentic, OpenManus) and released model families including GPT-4o, Claude, Qwen, and CodeLlama.
Only FC2-positive traces (failure modes 2.1--2.6) are candidates for conversion.

The pilot and AG2-scoping decisions use a locked Step~1 train/hold-out inventory of train FC2-positive $n{=}423$ (Table~\ref{tab:train-fc2-inventory}) and hold-out FC2-positive $n{=}209$ (Table~\ref{tab:holdout-fc2-inventory}) under an \textbf{AG2-by-benchmark} split: train units are AG2/GSM, AG2/Olympiad, ChatDev, MetaGPT, AppWorld, and HyperAgent; hold-out units are Magentic, OpenManus, and AG2/MMLU.
Hold-out status is inferred from that framework-based split; \texttt{MAD\_full\_dataset.json} contains no explicit train/test field.
The conversion pilot samples only from the train pool.
Table~\ref{tab:train-fc2-inventory} is the scoping inventory for the pilot and for AG2 exclusion; it is not the MetaGPT sampling frame for the later conversion pass.

The locked conversion pass then codes non-AG2 traces under the same AG2-by-benchmark train definition, but expands MetaGPT coding across the four released model families available in the working extract (GPT-4o, Claude, Qwen, CodeLlama).
Table~\ref{tab:conversion-accounting} is therefore the source of truth for conversion counts: MetaGPT contributes 217 coded traces there, not the Step~1 MetaGPT cell of 100. The two counts differ because Step~1 was scoped for the pilot sample rather than for full conversion: it drew MetaGPT traces from a single model family, whereas the conversion pass covers all four MetaGPT model families present in the working extract (the locally cached subset of \texttt{MAD\_full\_dataset.json} used throughout, released with the deposit). The MetaGPT FC2+ train pool in that extract totals 217 traces, all of which were coded.
AG2 GSM/Olympiad remains documentation-only \textsc{Structural Mismatch} under both tables.

\subsection{Convertibility criteria}

An FC2-labeled trace enters the ToMAS conversion pool only if \emph{all four} of the following hold: \textbf{(a)} a peer agent has an identifiable role, knowledge state, belief, or intention; \textbf{(b)} another agent acts without correctly accounting for that state; \textbf{(c)} the trace contains enough evidence to identify a better-coordinated action; and \textbf{(d)} a reasonably clear evaluation question and answer can be constructed.
Meeting the MAST FC2 label alone is not sufficient.
Traces that fail any criterion are coded \textsc{Ambiguous}, \textsc{Not Convertible}, or \textsc{Structural Mismatch} and never generate items or rewards.

\textsc{Structural Mismatch} applies when inclusion (a) fails---for example, AG2 traces pairing an \texttt{assistant} with a \texttt{mathproxyagent} that only executes code.
\textsc{Ambiguous} applies when a peer exists but multiple incompatible readings are equally supported, or a non-ToM co-cause (e.g., a framework \texttt{gui\_design} flag) blocks isolation of the partner-state question.

To assess the reliability of the four-category scheme, the first author and supervisor independently categorized all 18 sampled traces using the rubric defined above. Raw agreement was 94.4\% (17/18 items), yielding Cohen's $\kappa = 0.92$, indicating almost perfect agreement~\cite{landis1977measurement}. The single disagreement occurred on trace \#13 (MetaGPT), where the first author assigned AMBIGUOUS and the supervisor assigned NOT CONVERTIBLE; the supervisor's coding was adopted as the final label, yielding CLEAN: 4, AMBIGUOUS: 4, NOT CONVERTIBLE: 2, STRUCTURAL MISMATCH: 8.

Of the 423 train FC2+ traces in the locked Step~1 inventory (Table~\ref{tab:train-fc2-inventory}), the \textbf{258} AG2 GSM/Olympiad cases (127 GSM + 131 Olympiad) are scoped out of the convertible partner-state pool and retained only for documentation as \textsc{Structural Mismatch}: their FC2 labels reflect solver--tool-proxy protocol issues rather than an isolable peer belief or knowledge target (confirmed on all 8/8 AG2 traces in the reliability pilot).
The Step~1 non-AG2 remainder is \textbf{165} traces (ChatDev 53, MetaGPT 100, AppWorld 6, HyperAgent 6); that MetaGPT cell of 100 is the pilot-era inventory count, not the conversion count.
The locked conversion log records \textbf{242} coded non-AG2 traces (Table~\ref{tab:conversion-accounting}), of which MetaGPT accounts for \textbf{217} traces across four model families.
ChatDev coding was stopped after 13 GPT-4o traces because every coded ChatDev case was \textsc{Ambiguous} under the locked rubric (often \texttt{gui\_design} co-cause), so the remaining 40 ChatDev inventory traces were not converted into items.
AppWorld and HyperAgent were fully coded (6/6 each).
The coded pool yields \textbf{39} confirmed \textsc{Clean} items.
Category counts are taken from the locked conversion log (\texttt{conversion\_log.json}).

\begin{table}[t]
\caption{Locked Step~1 train FC2+ inventory used for the conversion pilot and AG2 scoping ($n{=}423$). This table does not supersede Table~\ref{tab:conversion-accounting} for MetaGPT conversion counts.}
\label{tab:train-fc2-inventory}
\centering
\scriptsize
\begin{tabular}{@{}lrr@{}}
\toprule
\textbf{Unit} & \textbf{Train FC2+} & \textbf{Role in this paper} \\
\midrule
AG2/GSM & 127 & Excluded before partner-state coding (structural mismatch) \\
AG2/Olympiad & 131 & Excluded before partner-state coding (structural mismatch) \\
ChatDev & 53 & Pilot inventory; coding later stopped after 13 GPT-4o traces \\
MetaGPT & 100 & Pilot inventory only; conversion uses Table~\ref{tab:conversion-accounting} (217) \\
AppWorld & 6 & Pilot inventory; all 6 later coded \\
HyperAgent & 6 & Pilot inventory; all 6 later coded \\
\midrule
\textbf{AG2 subtotal} & \textbf{258} & Documentation only \\
\textbf{Non-AG2 subtotal} & \textbf{165} & Pilot-era conversion pool \\
\textbf{Total} & \textbf{423} & Pilot / AG2-scoping lock \\
\bottomrule
\end{tabular}
\end{table}

\begin{table}[t]
\caption{Conversion-pass accounting by framework and model (authoritative for coded counts). AMB\,=\,\textsc{Ambiguous}; NC\,=\,\textsc{Not Convertible}; SM\,=\,\textsc{Structural Mismatch}. Coded is the row sum. MetaGPT $217=53{+}47{+}31{+}86$ supersedes the Step~1 MetaGPT cell of 100.}
\label{tab:conversion-accounting}
\centering
\scriptsize
\begin{tabular}{@{}llrrrrr@{}}
\toprule
\textbf{Framework} & \textbf{Model} & \textbf{Coded} & \textbf{CLEAN} & \textbf{AMB} & \textbf{NC} & \textbf{SM} \\
\midrule
MetaGPT & GPT-4o & 53 & 28 & 2 & 23 & 0 \\
MetaGPT & Claude & 47 & 4 & 0 & 43 & 0 \\
MetaGPT & Qwen & 31 & 4 & 2 & 25 & 0 \\
MetaGPT & CodeLlama & 86 & 0 & 2 & 48 & 36 \\
AppWorld & GPT-4o & 6 & 2 & 0 & 4 & 0 \\
HyperAgent & Claude & 6 & 1 & 2 & 3 & 0 \\
ChatDev & GPT-4o & 13 & 0 & 13 & 0 & 0 \\
\midrule
\textbf{Total} & & \textbf{242} & \textbf{39} & \textbf{21} & \textbf{146} & \textbf{36} \\
\bottomrule
\end{tabular}
\end{table}

\subsection{Item construction and reward}
\label{sec:reward}

Each \textsc{Clean} trace becomes one functional partner-state item.
The model receives a truncated coordination context (task text, roles, prior messages up to the miscoordinated action, visible tool outputs) but not the MAST failure label or later revealing turns.
The model must state the peer's relevant informational state and/or the better-coordinated next action; scoring is against fixed gold under a documented rubric sheet.

For \textsc{ToMAS-RL} training we use a single transparent reward: \textbf{binary ROUGE-L F1 match} on \textsc{Clean} training items ($+1$ if ROUGE-L F1 exceeds 0.2; $0$ otherwise).
Scores are computed with \texttt{rouge-score}'s \texttt{RougeScorer(["rougeL"])} default tokenizer on the extracted \texttt{[ANSWER]} span versus the fixed gold string; the reported value is \textbf{F1} (not precision or recall alone).
The training threshold $0.2$ is intentionally loose and was selected because short partner-state answers rarely reach high lexical overlap, and a stricter cutoff collapsed the binary reward toward all zeros on the 39-item pool in preliminary checks, removing the GRPO learning signal (0.2 selected for a usable training signal; held-out uses $0.4$).
Held-out evaluation uses the stricter $>0.4$ cutoff to reduce spurious lexical matches on longer Magentic/GAIA answers where partial n-gram overlap is common even when the partner-state claim is wrong.
Gold is assigned manually at construction; matching at train time is automatic.
Only \textsc{Clean} items enter the reward pool.

Circularity is controlled by source separation: rewards use train-split \textsc{Clean} items, while evaluation uses separate held-out \textsc{Clean} items.
Training prompts never include MAST failure tags or post-failure turns.
The present diagnostic scores generated answers against fixed held-out gold strings and is not an independent estimate of the MAST FC2 failure rate.

\subsection{Illustrative conversion examples}
\label{sec:conversion-examples}

Table~\ref{tab:conversion-examples} shows two \textsc{Clean} items and one \textsc{Ambiguous} case from the approved conversion batches, clarifying when a partner-state gap is isolable under criteria~(a)--(d).

\begin{table}[t]
\centering
\caption{Illustrative \textsc{Clean} and \textsc{Ambiguous} conversion examples.}
\label{tab:conversion-examples}
\begin{tabular}{@{}lp{0.72\columnwidth}@{}}
\toprule
\textbf{Label} & \textbf{Example} \\
\midrule
\textsc{Clean} A & \textbf{AppWorld \#15.}
\textit{Trace:} \texttt{AppWorld/Test-C/GPT-4o/18}.
\textit{Category:} \textsc{Clean}.
\textit{Context:} Supervisor agent issues \texttt{show\_account\_passwords}; Spotify agent has previously requested an access token.
\textit{Question:} What credential does the Spotify agent require, and what should the Supervisor provide instead?
\textit{Gold:} The Spotify agent requires an OAuth access token, not an account password; the Supervisor should call the appropriate token-retrieval action.
\textit{Rationale:} Peer (Spotify) knowledge state is isolable; Supervisor action ignores the prior token request (criteria~(a)--(d)).
\textit{Supporting quote:} ``access\_token'' in Spotify agent request. \\
\midrule
\textsc{Clean} B & \textbf{MetaGPT \#12.}
\textit{Trace:} \texttt{MetaGPT/ProgramDev/Claude/91}.
\textit{Category:} \textsc{Clean}.
\textit{Context:} Coder emits cron string with 0 in hour field using HH:MM format; Tester checks scheduling behavior.
\textit{Question:} What cron format does the Coder treat as valid for daily execution?
\textit{Gold:} The Coder treats 0 HH:MM as valid but the hour field requires an integer 0--23, not HH:MM; Tester should check field format before testing schedule behavior.
\textit{Rationale:} Coder's format belief is recoverable from the trace; Tester should condition on that state before schedule checks (criteria~(a)--(d)).
\textit{Supporting quote:} cron string in Coder output. \\
\midrule
\textsc{Ambiguous} & \textbf{ChatDev \#11.}
\textit{Trace:} \texttt{ChatDev/ProgramDev/GPT-4o/82}.
\textit{Category:} \textsc{Ambiguous}.
\textit{Context:} Customer requests CLI interface; Programmer implements tkinter GUI; \texttt{gui\_design} parameter also set.
\textit{Question / gold:} Not issued---co-cause blocks a single partner-state item.
\textit{Rationale:} FC2 peer-state gap exists (Programmer ignores CLI requirement) but \texttt{gui\_design} system parameter independently influences output, preventing clean isolation of the inter-agent belief failure. \\
\bottomrule
\end{tabular}
\end{table}

\subsection{Training conditions and evaluation protocol}

We evaluate three conditions: a policy trained with the ToMAS reward
(\textsc{ToMAS-RL}), a policy trained with adapted
LLM-Coordination items (\textsc{Coord-RL}), and the untrained base
checkpoint (\textsc{Base}). Both trained conditions use
\texttt{Qwen/Qwen2.5-1.5B-Instruct} (Hugging Face Hub default revision at download time; revision not pinned; adapter configs record \texttt{revision=null}), the same optimization budget,
and two random seeds (42 and 43). Each training step generates four
completions. Training \texttt{generation\_kwargs} set only \texttt{max\_new\_tokens};
temperature and top-$p$ were left at Hugging Face \texttt{GenerationConfig}
defaults ($\mathrm{temperature}{=}1.0$, $\mathrm{top\_p}{=}1.0$). Held-out evaluation uses greedy decoding
(\texttt{do\_sample=False}), so temperature and top-$p$ are unused at test time.

We use Group Relative Policy Optimization (GRPO) with the binary
reward defined in Section~\ref{sec:reward}. We selected the 1.5B
Instruct checkpoint because it supports LoRA training and GRPO group
rollouts on a Colab T4 without gated access. We use GRPO rather than
dynamic sampling because the confirmed \textsc{Clean} pool contains
only 39 items; removing all-correct or all-incorrect groups could
further reduce an already limited training signal.

Table~\ref{tab:train-config} summarizes the shared GRPO training configuration.
Each step samples one item from the 39-item training pool, generates 4 completions (batch $=$ 1 prompt $\times$ 4 completions), and applies binary ROUGE-L reward.
Over 200 steps, each item is presented approximately $200/39 \approx 5.1$ times on average (one ``pass'' $=$ one presentation of that item).
%We do not report ``1.28 epochs'': that figure equals $200/(39{\times}4)$ and incorrectly treats the four GRPO completions as distinct dataset items.

\begin{table}[t]
\centering
\caption{Shared GRPO training configuration.}
\label{tab:train-config}
\scriptsize
\begin{tabular}{@{}lp{0.58\columnwidth}@{}}
\toprule
\textbf{Parameter} & \textbf{Value} \\
\midrule
Base model & \texttt{Qwen/Qwen2.5-1.5B-Instruct} \\
Hub revision & Default at download (not pinned; \texttt{revision=null}) \\
Libraries & \texttt{trl} 1.12.0; \texttt{transformers} 5.15.1; \texttt{peft} 0.20.0; \texttt{torch} 2.11.0+cu128 \\
LoRA & rank 8, alpha 16, dropout 0; targets \texttt{q\_proj}, \texttt{v\_proj} \\
RL / optimizer & GRPO (\texttt{trl.GRPOTrainer}); AdamW (\texttt{trl} default) \\
LR schedule & TrainingArguments defaults (linear; \texttt{warmup\_ratio}=0); LR $1{\times}10^{-6}$ \\
Steps / batch & 200 steps; per step: 1 prompt $\times$ 4 completions (GRPO group size 4); grad.\ accum.\ 1 \\
Lengths / stop & Prompt 512; completion 128; stop \texttt{[/ANSWER]} \\
Train sampling & Temperature $1.0$, top-$p$ $1.0$ (HF defaults; no override) \\
Held-out decoding & Greedy (\texttt{do\_sample=False}) \\
Reward & ROUGE-L \textbf{F1}; train $>0.2$ (0.2 selected); held-out $>0.4$ \\
Sampling schedule & 1 item/step $\times$ 200 steps; $\approx$5.1 presentations/item ($200/39$) \\
Seeds / hardware & 42, 43; Colab T4 ${\sim}$3.5GB; runtime not archived \\
Code / logs & \texttt{train\_condition.py}; \texttt{checkpoint-200}; \texttt{logs/*\_rewards.jsonl} \\
\bottomrule
\end{tabular}
\end{table}

Each trained condition runs for 200 optimization steps. The
\textsc{ToMAS-RL} condition trains on the 39 converted items.
The \textsc{Coord-RL} condition uses an equal number of adapted
LLM-Coordination items under the same binary reward format and
training budget.
Adaptation procedure: from the public LLM-Coordination CoordQA CSVs
(\texttt{OvercookedFixedHeadings.csv}, \texttt{HanabiFixedHeadings.csv},
\texttt{CollabGamesFixedHeadings.csv}), we retained Theory-of-Mind
questions only (excluding Environment Comprehension and Joint Planning),
yielding 66 eligible rows; we then sampled 39 items with seed~42
(Overcooked 15, Hanabi 14, Collab Capture 6, Collab Escape 4).
Each row was mapped to the ToMAS JSONL schema
(\texttt{id}, \texttt{source}, \texttt{prompt}, \texttt{question},
\texttt{gold}, \texttt{supporting\_quote}), with scenario/state text as
\texttt{prompt}, the ToM question as \texttt{question}, and the published
ToM answer as \texttt{gold}, then scored with the same binary ROUGE-L
reward as \textsc{ToMAS-RL}.
The held-out diagnostic contains 28 Magentic GAIA
items. One observation is one generated response to one held-out
item.

\begin{table}[t]
\caption{Conditions retained in the feasibility experiment.}
\label{tab:conditions}
\centering
\begin{tabular}{@{}lp{0.60\columnwidth}@{}}
\toprule
\textbf{Condition} & \textbf{Signal} \\
\midrule
\textsc{ToMAS-RL}
    & Binary partner-state reward on 39 converted
      \textsc{Clean} training items \\

\textsc{Coord-RL}
    & 39 LLM-Coordination ToM items (seed~42 subsample of 66);
      mapped to ToMAS JSONL; matched reward format and budget \\

\textsc{Base}
    & Untrained Qwen2.5-1.5B-Instruct checkpoint \\
\bottomrule
\end{tabular}
\end{table}

We considered EnactToM as a functional Theory-of-Mind baseline.
However, its hard items require PDDL-predicate JSON responses, whereas
the retained conditions use natural-language answers and ROUGE-L
scoring. Because adapting these formats would introduce an additional
methodological difference, EnactToM was excluded before the
comparison. Consequently, the present experiment does not include a
direct functional Theory-of-Mind baseline.

The diagnostic reports binary ROUGE-L outcomes rather than an
independent estimate of the MAST FC2 failure rate. Given the small
held-out pool and the domain mismatch described below, we treat the
results descriptively and do not perform inferential comparisons.

\FloatBarrier
\section{Results}

We report findings in relation to RQ1 (convertibility) and RQ2 (small-scale GRPO training diagnostic / cross-domain evaluation).

\subsection{Dataset inventory}

The Step~1 lock reports 423 train FC2+ (Table~\ref{tab:train-fc2-inventory}) and 209 hold-out FC2+ traces (Table~\ref{tab:holdout-fc2-inventory}: Magentic 103, OpenManus 6, AG2/MMLU 100) under the AG2-by-benchmark split.
That inventory supports the pilot sample and AG2 scoping; conversion outcomes follow Table~\ref{tab:conversion-accounting}.
Neither inventory is an RL outcome.

\begin{table}[t]
\caption{Locked Step~1 hold-out FC2+ inventory under the same AG2-by-benchmark split ($n{=}209$). The held-out diagnostic uses only the Magentic slice.}
\label{tab:holdout-fc2-inventory}
\centering
\scriptsize
\begin{tabular}{@{}lrr@{}}
\toprule
\textbf{Unit} & \textbf{Hold-out FC2+} & \textbf{Role in this paper} \\
\midrule
Magentic & 103 & Source pool for the 28-item diagnostic \\
OpenManus & 6 & Not used in the present diagnostic \\
AG2/MMLU & 100 & Not used in the present diagnostic \\
\midrule
\textbf{Total} & \textbf{209} & Step~1 hold-out lock \\
\bottomrule
\end{tabular}
\end{table}

\subsection{Conversion pilot}

From the train FC2-positive traces we drew a stratified random sample of $n{=}18$ ($\mathrm{seed}{=}42$).
Each trace was coded \textsc{Clean}, \textsc{Ambiguous}, \textsc{Not Convertible}, or \textsc{Structural Mismatch} under the criteria in Section~III.

\begin{table}[t]
\caption{Conversion-pilot outcomes on $n{=}18$ train FC2-positive traces.}
\label{tab:pilot}
\centering
\begin{tabular}{@{}lrrr@{}}
\toprule
\textbf{Code} & \textbf{AG2} & \textbf{Non-AG2} & \textbf{Total} \\
 & \textbf{($n{=}8$)} & \textbf{($n{=}10$)} & \textbf{($n{=}18$)} \\
\midrule
CLEAN & 0 & 4 & 4 \\
AMBIGUOUS & 0 & 4 & 4 \\
NOT CONVERTIBLE & 0 & 2 & 2 \\
STRUCTURAL MISMATCH & 8 & 0 & 8 \\
\bottomrule
\end{tabular}
\end{table}

Table~\ref{tab:pilot} reports the measured codes.
All \textbf{8/8} sampled AG2 GSM/Olympiad traces were coded \textsc{Structural Mismatch}---labeled FC2 in those traces reflects solver--tool-proxy protocol issues rather than peer-belief reasoning in this small sample.
We treat that pilot AG2 pattern as confirmed for scoping: the AG2 GSM/Olympiad train FC2+ traces ($n{=}258$ in the locked inventory, including the 8 AG2 pilot samples) are retained for documentation as \textsc{Structural Mismatch}, not as convertible partner-state items, and are excluded from the partner-state conversion pool.
Among \textbf{10 non-AG2} samples, \textbf{4 were CLEAN (40\%)}, 4 \textsc{Ambiguous} (including a ChatDev \texttt{gui\_design} co-cause), and 2 \textsc{Not Convertible}.

A conversion pass recorded \textbf{242} coded non-AG2 traces in the locked log (Table~\ref{tab:conversion-accounting}; MetaGPT 217 across four model families) and yielded \textbf{39 confirmed CLEAN items} with further yield limited under the locked rubric.
CLEAN rates differed sharply by model within MetaGPT: GPT-4o $\approx$53\% (28/53), Qwen $\approx$13\% (4/31), Claude $\approx$9\% (4/47), and CodeLlama 0\% (0/86); ChatDev was 0\% CLEAN across 13 coded traces (all \textsc{Ambiguous}).
These descriptive rates (Table~\ref{tab:conversion-accounting}) show that recoverable partner-state items
were substantially more common in the available GPT-4o MetaGPT
traces than in the other coded sources. However, they should not be
interpreted as causal evidence that one model has stronger
partner-state reasoning than another because the trace contents,
tasks, and failure distributions were not experimentally controlled.
Qualitatively, many non-convertible CodeLlama and ChatDev traces were
dominated by prompt echo, repetitive question--answer loops, or
framework-level configuration conflicts rather than an isolable
peer-state reasoning failure.

Two confirmed \textsc{Clean} drafts illustrate the item format (AppWorld token confusion; MetaGPT crontab field misuse).
A ChatDev CLI-vs-GUI case is retained as an \textsc{Ambiguous} illustration because framework-injected GUI configuration co-causes the drift.
The pilot establishes that \textbf{FC2 label $\neq$ ToMAS-convertible} under our criteria.

\subsection{Training and held-out evaluation}

\begin{table}[t]
\caption{Training reward at selected optimizer steps. Each cell is the \texttt{trl}-logged \texttt{rewards/binary\_reward/mean} written by the reward callback at that step (logs every 10 steps; \texttt{trl} 1.12.0). Values equal exact small-denominator ratios recovered from \texttt{experiments/logs/*\_rewards.jsonl} (e.g., $0.417{=}5/12$, $0.833{=}5/6$, $0.125{=}1/8$), so they are \emph{not} means over a fixed 40-sample (4$\times$10) window. Seeds 42 and 43 use different random initializations.}
\label{tab:training-rewards}
\centering
\begin{tabular}{@{}lrrrrr@{}}
\toprule
\textbf{Condition} & \textbf{Seed} & \textbf{Step 50} & \textbf{Step 100} & \textbf{Step 150} & \textbf{Step 200} \\
\midrule
\textsc{ToMAS-RL} & 42 & 0.417 & 0.750 & 0.833 & 0.125 \\
\textsc{ToMAS-RL} & 43 & 0.417 & 0.500 & 0.167 & 0.875 \\
\textsc{Coord-RL} & 42 & 0.667 & 0.250 & 0.583 & 0.250 \\
\textsc{Coord-RL} & 43 & 0.750 & 0.625 & 0.167 & 0.500 \\
\bottomrule
\end{tabular}
\end{table}

All trained conditions produced nonzero logged rewards throughout training (Table~\ref{tab:training-rewards}), confirming that the reward function fired and the GRPO loop ran to completion.

\paragraph{RQ2 adapter checks}\label{sec:adapter-checks} Identical greedy held-out answers across \textsc{ToMAS-RL}, \textsc{Coord-RL}, and \textsc{Base} (Tables~\ref{tab:heldout-item-hits}--\ref{tab:heldout-positive-answers}) prompted three checks.
(a)~Evaluation attaches adapters via \texttt{PeftModel.from\_pretrained} on each \texttt{checkpoint-200} directory (\texttt{experiments/evaluate.py}); logs confirm ``Attached adapter'' for all four trained runs.
(b)~On the 39 training items under the same greedy protocol, \textsc{Coord-RL} matched \textsc{Base} on $39/39$ items (both seeds).
\textsc{ToMAS-RL} matched on $37/39$ (seed~42) and $38/39$ (seed~43); the few mismatches are minor surface paraphrases on the same items (\texttt{item\_34}, and \texttt{item\_35} for seed~42 only), not systematic answer changes.
(c)~Adapter norms: LoRA~$A$ dominates ($\mathrm{L2}\!\approx\!12.22$, typical of nonzero init), while LoRA~$B$ remains near zero ($\mathrm{L2}\!\approx\!0.005$--$0.006$, $\max|w|\!\approx\!4{\times}10^{-5}$); with $\alpha/r{=}2$, effective $\max|\Delta W|\!\approx\!7{\times}10^{-6}$.
Plainly: adapters were loaded and training ran, but under LR~$1{\times}10^{-6}$ the policy barely moved---held-out outputs are identical to \textsc{Base}, and training-set outputs are identical or near-identical---so RQ2 does not establish a coordination-training effect.

Held-out items were drawn from the Magentic slice of the locked hold-out FC2+ pool (Table~\ref{tab:holdout-fc2-inventory}: Magentic 103, OpenManus 6, AG2/MMLU 100; total 209).
OpenManus and AG2/MMLU were not used for this diagnostic.
The first author applied the same locked convertibility rubric used in training conversion and retained traces that yielded isolable partner-state items with recoverable gold, producing the released set of 28 \textsc{Clean} Magentic/GAIA items (\texttt{IVH\_01}--\texttt{IVH\_28}).
Gold answers for all 28 held-out items were created by the first author using the same locked convertibility rubric applied to the training conversion; no second annotator re-checked held-out gold for this pilot.
Per-item conversion rationales for held-out items follow the locked rubric and are included in the Zenodo release alongside truncated context, question, and gold.
No held-out items overlap with training items by source trace ID (verified by ID intersection check).
We did not run a blinded human or semantic-equivalence assessment in this pilot; the reported held-out metric is therefore lexical ROUGE-L F1 only, which can credit phrasing overlap without verifying partner-state reasoning.

\begin{table}[t]
\caption{Held-out items with binary ROUGE-L $>0.4$ by condition and seed ($n{=}28$ items each). Positive item IDs and generated answers are identical across \textsc{ToMAS-RL}, \textsc{Coord-RL}, and \textsc{Base} under greedy decoding (\texttt{do\_sample=False}). Answers are the extracted \texttt{[ANSWER]} spans from the evaluation runs summarized here; the same two items fire under every condition.}
\label{tab:heldout-item-hits}
\centering
\scriptsize
\begin{tabular}{@{}llrl@{}}
\toprule
\textbf{Condition} & \textbf{Seed} & \textbf{$n{>}0.4$} & \textbf{Positive item IDs} \\
\midrule
\textsc{ToMAS-RL} & 42 & 2 & IVH\_04, IVH\_12 \\
\textsc{ToMAS-RL} & 43 & 2 & IVH\_04, IVH\_12 \\
\textsc{Coord-RL} & 42 & 2 & IVH\_04, IVH\_12 \\
\textsc{Coord-RL} & 43 & 2 & IVH\_04, IVH\_12 \\
\textsc{Base} & --- & 2 & IVH\_04, IVH\_12 \\
\bottomrule
\end{tabular}
\end{table}

\begin{table}[t]
\caption{Generated answers for the two threshold-passing held-out items (identical under every condition/seed in Table~\ref{tab:heldout-item-hits}).}
\label{tab:heldout-positive-answers}
\centering
\scriptsize
\begin{tabular}{@{}llp{0.52\columnwidth}@{}}
\toprule
\textbf{Item} & \textbf{ROUGE-L} & \textbf{Generated answer} \\
\midrule
IVH\_04 & 0.404 & WebSurfer knows 2072 Akaikai Loop sold for \$1{,}190{,}000 (14 Dec 2022) while 2017 Komo Mai Drive sold for \$1{,}080{,}000 (17 Aug 2023); the acting agent should report which 2022 sale was higher. \\
IVH\_12 & 0.476 & Peers establish honey density 1.38--1.45\,kg/L (20$^{\circ}$C) and mayonnaise 0.91\,kg/L; the acting agent should compute cups of honey to remove so honey weighs less than mayonnaise. \\
\bottomrule
\end{tabular}
\end{table}

Because every condition decodes identically to \textsc{Base} (Tables~\ref{tab:heldout-item-hits} and~\ref{tab:heldout-positive-answers}), the held-out numbers describe base-model behaviour on the diagnostic and carry no information about training.
We report them for completeness and because they expose a second limitation independent of the optimization failure: a provenance gap between the two item pools.
Training items are drawn predominantly from MetaGPT code-coordination failures (36/39 items, with 2 AppWorld and 1 HyperAgent), whereas held-out items are general-purpose GAIA tasks.
This asymmetry follows from the AG2-by-benchmark split fixed at Step~1 and was therefore a property of the design, not a result of the experiment.
The two threshold-passing items also illustrate the scoring limitation: both clear $0.4$ on phrasing overlap with gold rather than on a verified partner-state claim.
Since every condition and seed produced the same two item IDs and the same extracted answers under greedy decoding, we report the compact hit table rather than a redundant $28{\times}$condition matrix; non-hitting items scored $\leq 0.4$ throughout.
A conclusive study requires a matched-domain held-out pool, an effective learning rate, and semantic or human scoring.

\FloatBarrier
\section{Discussion}

The conversion pilot and full conversion pass show that a nonempty convertible subset of FC2 can be represented as functional partner-state problems, and that the same items are well-formed as binary rewards.
They also tied the claim: convertibility must be filtered, not assumed from the FC2 label alone.
The AG2 GSM/Olympiad pattern is treated as \textsc{Structural Mismatch} for ToMAS construction; ChatDev and CodeLlama MetaGPT are excluded after yielding 0\% CLEAN under the locked rubric.
Model-level convertibility (highest for GPT-4o MetaGPT; near-zero for CodeLlama and ChatDev) is reported as a finding about which execution traces support partner-state items, rather than merely a limit on pool size.

The present 28-item diagnostic does not establish a training effect (RQ2): adapters load at evaluation, LoRA~$\Delta W$ remains near zero under LR~$1{\times}10^{-6}$, held-out greedy outputs match \textsc{Base}, and training-set greedy outputs match on $37$--$39/39$ items (only minor paraphrases elsewhere).
Convertibility findings address RQ1.
Two further limitations would have blocked a conclusive result even had the policy updated: the provenance gap between MetaGPT-derived training items and GAIA-derived held-out items, and scoring by lexical overlap alone.
Cross-framework generalization therefore remains open.
We note also that a positive result under a corrected setup would show that ToMAS items are trainable, not that failure grounding \emph{per se} is what produced the gain; isolating that would require a matched control trained on authored items of equivalent difficulty.
Convertibility analysis under RQ1 is limited by cell size in the 39 confirmed CLEAN items from the conversion pass.

\section{Conclusion}

Our study highlights how systematic diagnosis of agent errors can serve as a direct source for improving social reasoning in large language models. ToMAS demonstrates that a subset of MAST FC2 failures can support a failure-grounded Theory of Mind benchmark under a structured conversion rubric with near-perfect inter-annotator agreement ($\kappa{=}0.92$). Convertibility varies markedly with the structure of the source model's output. The training diagnostic did not produce a measurable policy update and therefore yields no evidence about coordination training; together with the item-provenance gap and lexical scoring, it defines what a conclusive follow-up study must control.

\section*{Data Availability}
ToMAS evaluation items, annotation materials, scoring scripts, \texttt{train\_condition.py}, GRPO reward logs, and Coord-RL training items are available at \url{https://doi.org/10.5281/zenodo.22685705}~\cite{ishfaq2026tomaszenodo} under Creative Commons Attribution 4.0 International (CC-BY-4.0). Items derived from MAST-Data must also attribute \texttt{mcemri/MAST-Data}~\cite{cemri2025mast}.
\FloatBarrier
\bibliographystyle{IEEEtran}
\bibliography{main}

\end{document}